\pdfoutput=1
\documentclass[reprint,aps,prd,superscriptaddress,showpacs,preprintnumbers,amsmath,amssymb,floatfix]{revtex4-2} 

\usepackage[euler]{textgreek}
\usepackage[mathlines]{lineno}
\usepackage{graphicx,subfigure}
\usepackage{epstopdf}
\usepackage{dcolumn}
\usepackage{bm}
\usepackage{color}
\usepackage{upgreek}
\usepackage{afterpage}
\usepackage{placeins}
\usepackage{tikz, pgfplots}
\usepackage{mhchem} 
\usepackage[colorlinks=true]{hyperref}
\usepackage[user,titleref]{zref}

\newcommand{\keVee}{keV$_{\text{ee}}$} 
\newcommand{\eVee}{eV$_{\text{ee}}$} 

\newcommand{\mus}{\ensuremath{\mu{}s}} 

\definecolor{kittycolor}{RGB}{168,168,244}

\begin{document}

\preprint{ \today}

\title{First observation of isolated nuclear recoils following neutron capture for dark matter calibration}

\author{A.N.~Villano}\email{Corresponding author: anthony.villano@ucdenver.edu} \affiliation{Department of Physics, University of Colorado Denver, Denver, Colorado 80217, USA}
\author{M.~Fritts} \affiliation{School of Physics \& Astronomy, University of Minnesota, Minneapolis, Minnesota 55455, USA}
\author{N.~Mast} \affiliation{School of Physics \& Astronomy, University of Minnesota, Minneapolis, Minnesota 55455, USA}
\author{S.~Brown} \affiliation{Department of Physics, University of Colorado Denver, Denver, Colorado 80217, USA} 
\author{P.~Cushman} \affiliation{School of Physics \& Astronomy, University of Minnesota, Minneapolis, Minnesota 55455, USA}
\author{K.~Harris}  \affiliation{Department of Physics, University of Colorado Denver, Denver, Colorado 80217, USA}
\author{V.~Mandic} \affiliation{School of Physics \& Astronomy, University of Minnesota, Minneapolis, Minnesota 55455, USA}

\smallskip
\date{\today}

\noaffiliation


\smallskip

\begin{abstract}
Low-energy nuclear recoils (NRs) are hard to measure because they produce few e$^{-}$/h$^+$ pairs
in solids -- i.e.~they have low ``ionization yield.'' A silicon detector was exposed to thermal
neutrons over 2.5\,live-days, probing NRs down to 450\,eV. The observation of a neutron
capture-induced component of NRs at low energies is supported by the much-improved fit upon
inclusion of a capture NR model. This result shows that thermal neutron calibration of very low
recoil energy NRs is promising for dark matter searches, coherent neutrino experiments, and
improving understanding of ionization dynamics in solids.

\end{abstract}


\maketitle

%
%
%
%
%
%
%

%
%
%
%
%
%
%
%

\section{\label{sec:intro}Introduction}
The observation of 100\,eV-scale nuclear recoils (NRs) is a decades-long detector challenge that
is only recently becoming accessible due to new technological
advances~\cite{PhysRevLett.122.161801,PhysRevD.102.091101,PhysRevLett.127.061801,PhysRevLett.123.251801,PhysRevD.100.102002,PhysRevD.101.042001}.
While the theoretical framework remains deeply rooted in work from the
1960s~\cite{osti_4701226,Izraelevitch_2017,PhysRevD.94.082007,PhysRevD.94.122003,PhysRevD.91.083509,PhysRevD.42.3211,PhysRevA.45.2104}\footnote{The
first reference here is the Lindhard model of the 1960's and the rest of the papers are
experimental studies spanning almost three decades and beginning almost thirty years after the
Lindhard model's publication and yet they all make exclusive use of that model and mention it
\emph{in the abstract or introduction} despite it being known to be flawed for low energies.}, a
better and more modern understanding of these low-energy recoils is crucial for progress in
several contemporary fundamental physics fields,  including dark matter (DM) direct detection and
Coherent Elastic Neutrino-Nucleus Scattering (CE$\nu$NS). We have observed isolated NRs in this
energy region generated by the neutron capture process in silicon.  These NRs are not contaminated
by energy deposited by the outgoing gammas from the capture process and their recoil energies are
near threshold for even the most sensitive modern detectors. This technique has been recently
suggested~\cite{Thulliez_2021} but we believe this is the first observation of this kind, enabling
more detailed characterization studies of low-energy NRs.

The ultimate goal for this type of measurement is to use the exiting gammas for a coincidence tag
to make a high-precision measurement.  We have not used  this tagging in the present work but
have shown that even without the tagging the technique can be used to assess the NR detector
response -- including \emph{in situ} with low-background experiments. The present measurement has
key differences from previous measurements that utilize the capture
processes~\cite{PhysRevA.11.1347,PhysRevD.103.122003}. Those previous measurements of the neutron
capture allowed experimenters to observe an NR \emph{summed together with} 68\,keV of
electron-recoil (ER) energy -- these energy random variables may be correlated so that their
statistics are different when in each others' presence. In any case the NR energy is less than 1\%
of the total and relatively small fluctuations in the ER signal can have a large impact. 

\section{\label{sec:setup}Experimental Configuration}

NRs were detected in a silicon detector operated at cryogenic temperatures, specifically a
prototype SuperCDMS SNOLAB HV detector~\cite{PhysRevD.95.082002} read out with the SQUIDs and cold
hardware~\cite{AKERIB2008476} from CDMS-II Soudan, but modified to account for the lower
normal-state resistance of the new SuperCDMS SNOLAB HV transition-edge sensors
(TESs)~\cite{doi:10.1063/1.1146105}. The detector has a diameter of 100\,mm and a thickness of
33\,mm.  Each side has six phonon channels and each channel of the detector uses parallel arrays
of TESs to sense the phonon signal in the silicon substrate. The detector was mounted inside an
Oxford Instruments Kelvinox 100 dilution refrigerator~\cite{Kelvinox} at the University of
Minnesota and cooled to $\sim$30\,mK. It was  operated in the ``CDMSlite'' mode developed by the
SuperCDMS collaboration~\cite{PhysRevLett.112.041302}. A bias of -125\,V was used and six phonon
channels on one side were read out by prototype SuperCDMS detector control and readout boards
(DCRCs) at a 1.25\,MHz sampling rate~\cite{5874000}. 

\begin{figure}[]
    \centering
    \includegraphics[width=\columnwidth]{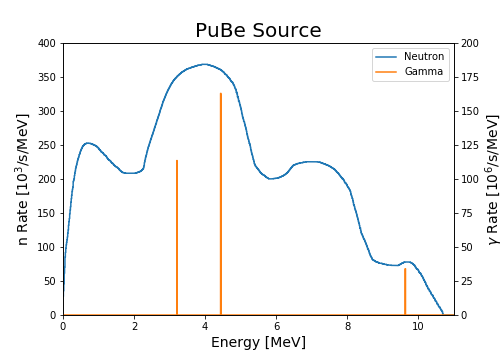}
    \caption{(Color online) PuBe source spectrum. The neutrons from the source were measured in
the early reference~\cite{PhysRev.98.740} and the gammas come from the accompanying de-excitation of the residual $^{12}$C system. The neutrons are emitted in broad groups corresponding to the excitation level of the residual system and the amount of stopping experience by the $\alpha$ prior to the ($\alpha$,n) reaction.}
    \label{fig:PuBe}
\end{figure}

The detector, when operated at high bias voltage, takes advantage of the Neganov, Trofimov, Luke
(NTL) effect for phonon amplification~\cite{doi:10.1063/1.341976,Neganov:1985khw}, in which the
phonon energy $E_t$ produced from a recoil of energy $E_r$ is dominated by secondary NTL phonons:
\begin{equation}
\label{eq:Etot}
    E_{t}=E_r\left(1+Y(E_r)\frac{e V}{\varepsilon_\gamma}\right),
\end{equation}
where $V$ is the bias voltage and $\varepsilon_\gamma$ is the average ER energy required to
produce an e/h pair (3.8 eV in Si~\cite{PEHL196845}). $Y(E_r)$ is a dimensionless quantity known
as the ionization yield and is normalized to unity for the mean ER response. The ionization from a
NR is less than half of that from an ER, and varies with recoil energy. Our detector calibrations
are based on an ER source so we refer to this energy scale as ``electron equivalent'' and denote
it by \eVee. The ionization yield determines how NRs appear on this energy scale.  
\begin{figure}[]
    \centering
    \includegraphics[width=\columnwidth]{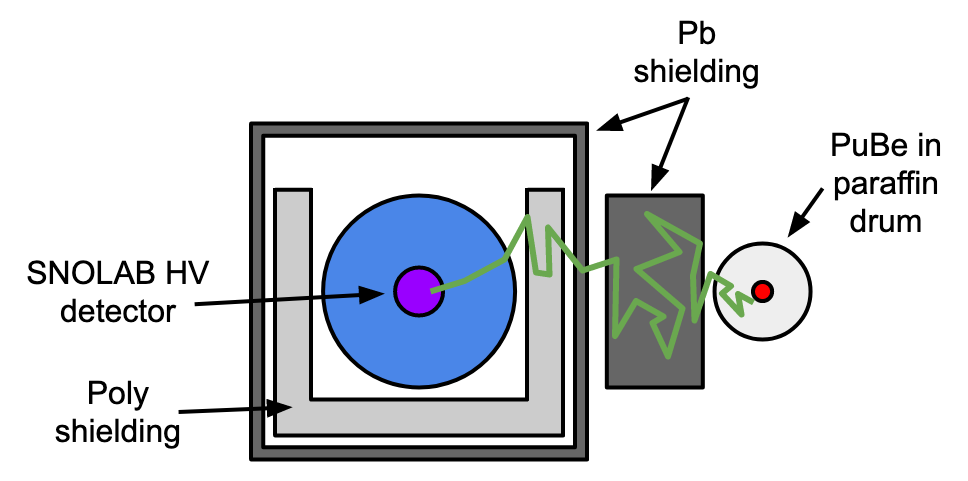}
	\caption{(Color online) Top-down view of experimental setup. Cartoon of thermalizing neutron shown in green.}
    \label{fig:R68_setup}
\end{figure}
Figure~\ref{fig:R68_setup} shows the experimental setup. Neutrons were produced by a PuBe source
(1.4\,Ci \textalpha, 62\,\textmu Ci n) enclosed in a paraffin-filled drum to reduce their energy.
The alpha and neutron rates are calculated based on the original source documentation, taking into
account changes over time from the decay of $^{241}$Pu to $^{241}$Am. See Fig.~\ref{fig:PuBe} for
the distribution of neutrons and gammas coming from the source.  The cryostat was shielded on
three sides and below by 20.3\,cm of polyethylene for further neutron moderation. It was also
surrounded on four sides by 1.6\,cm thick lead to reduce gamma backgrounds. Finally, a 30.5\,cm
lead wall was constructed to block direct $\sim$MeV gammas from the PuBe source. The wall was
placed near the PuBe drum in the line of sight to the SuperCDMS detector. A 1\,\textmu Ci
$^{241}\mathrm{Am}$ calibration source was mounted in the detector housing. The source
encapsulation effectively blocked gamma emission at energies below 5\,keV. A 1.6\,mm thick lead
disk with a 0.5\,mm diameter hole collimated the source gammas and restricted the emission rate to
less than 25\,Hz. A strip of Kapton tape placed over the collimator blocked alpha emission. 

\emph{Triggering.} A simple trigger quantity was defined as the difference between the average
sample value in two consecutive windows, the first 12.8\,$\mu$s wide and the second 4\,$\mu$s
wide.  If trigger thresholds were exceeded on any one of three phonon channels, a 3.2768\,ms trace
was recorded.  The trigger threshold values were set as low as possible while keeping the rate of
noise triggers below 150\,Hz.

Three datasets were taken for this study: one signal dataset with the PuBe source in place; a
background dataset with no external source; and a calibration dataset with a strong $^{22}$Na source
outside the cryostat. The total livetime for the PuBe dataset after cuts (see
Sec.~\ref{sec:analysis}) was approximately 2.5\,live-days.

\FloatBarrier
\section{\label{sec:analysis}Data analysis}
Phonon pulse amplitudes were extracted from raw traces using the Optimal Filter (OF) algorithm
(Appendix B of~\cite{Golwala_thesis}), which fits a pulse template to the measured trace by
minimizing the frequency-domain $\chi^2$ weighted by the measured noise spectrum. The algorithm
returns the best fit amplitude and start time. To decrease computational time, the start time is
required to fall within a 100\,\mus~window around the trigger time.

\subsection{Energy Calibration}
\label{sec:calibration}

\begin{figure}[]
    \centering
    \includegraphics[width=\columnwidth]{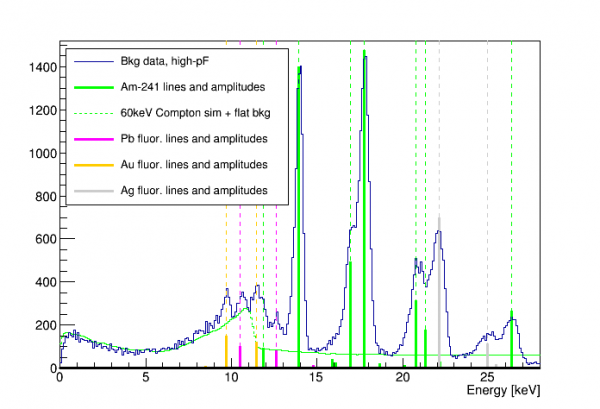}
    \caption{(Color online) Calibration lines in the background energy spectrum with selection of
events near the Am-241 calibration source.  }
\label{fig:spec_bknd}
\end{figure}

Energy calibration for ERs was performed using several low energy x-ray lines associated with the
$^{241}$Am source shown in Fig.~\ref{fig:spec_bknd}. The analysis range for our PuBe dataset is
50\,\eVee to 2\,\keVee.  The ER scale was calibrated separately for each dataset using two
prominent x-ray lines from $^{241}$Am: 14.0\,keV and 17.7\,keV. The data was fitted by assuming
that the OF zero corresponded to zero energy and employed a quadratic fit. Fits were nearly linear
with a small quadratic correction, accounting for TES saturation.  This fit showed good agreement
with the other five identified lines below 20\,keV down to our lowest line at 8\,keV (copper
fluorescence line that we can barely identify). The other identified fluorescence lines make sense
because our Am source was removed from a smoke detector where Pb, Au, and Ag are used in the
construction~\cite{osti_5999867}.  The lines came from Pb (10.5\,keV and 12.5\,keV), Au (9.6\,keV
and 11.5\,keV), and Ag (two above 20\,keV). These lines are identified in
Fig.~\ref{fig:spec_bknd}.

The energy resolution in \eVee was modeled as $\sigma_{\mathrm{ee}}^2 = \sigma_0^2 +
(B+\varepsilon_\gamma F)E_{\mathrm{ee}}$ and was fit using all the following lines: 9.6\,keV,
10.5\,keV, 12.5\,keV, 14.0\,keV, 17.7\,keV, and 22.1\,keV. A point at 0\,keV was included by using
the width of randomly-triggered noise event fits. The parameter $\sigma_0$ is the baseline
resolution, $F$=0.1161 is the ER Fano factor~\cite{PhysRev.72.26,LOWE1997354} (known to be
different for NRs), and $B$ incorporates any additional energy-dependent resolution effects
intrinsic to the detector. The widths of calibration lines and the baseline trace width determine
the best-fit values of $\sigma_0$=10$\pm$5\,\eVee~and $B$=1.9$\pm$0.1\,\eVee, with a $\chi^2$ of
$20.7$ for $5$ degrees of freedom. For ERs, detector and electronic effects have a stronger
impact on resolution than the Fano factor.

The low-energy ER calibration is of fundamental importance to our measurement because the NR
ionization is measured relative to it. We believe our calibration procedure -- outlined
above -- produces accurate ER energy measurements down to 50\,\eVee. Our nearly linear quadratic fit
outlined above for the OF energy tracks the integral of our phonon pulses linearly to our analysis
threshold even though it was only directly compared to a known line at 8\,keV. That means there is
no inherent bias in the OF above our analysis threshold. Furthermore, our assumption of the OF
zero corresponding to ``zero energy'' must be approximately valid because bias in the OF amplitude
is negligible compared to our trigger threshold (around 7\,\eVee) and contributes even less at our
analysis threshold (around 50\,\eVee). Between the zero point and our 8\,keV verification the only
plausible possibility is a monotonic calibration function; our nearly-linear function fits all
known evidence. When applying this procedure to a germanium detector of similar design and
operated at a similar voltage a good fit is obtained down to at least 100\,\eVee with direct
verification from a $^{71}$Ge electron capture line.

\subsection{Data Quality Cuts and Efficiencies}
\label{sec:cuts}

\begin{figure}[h]
    \centering
    \includegraphics[width=\columnwidth]{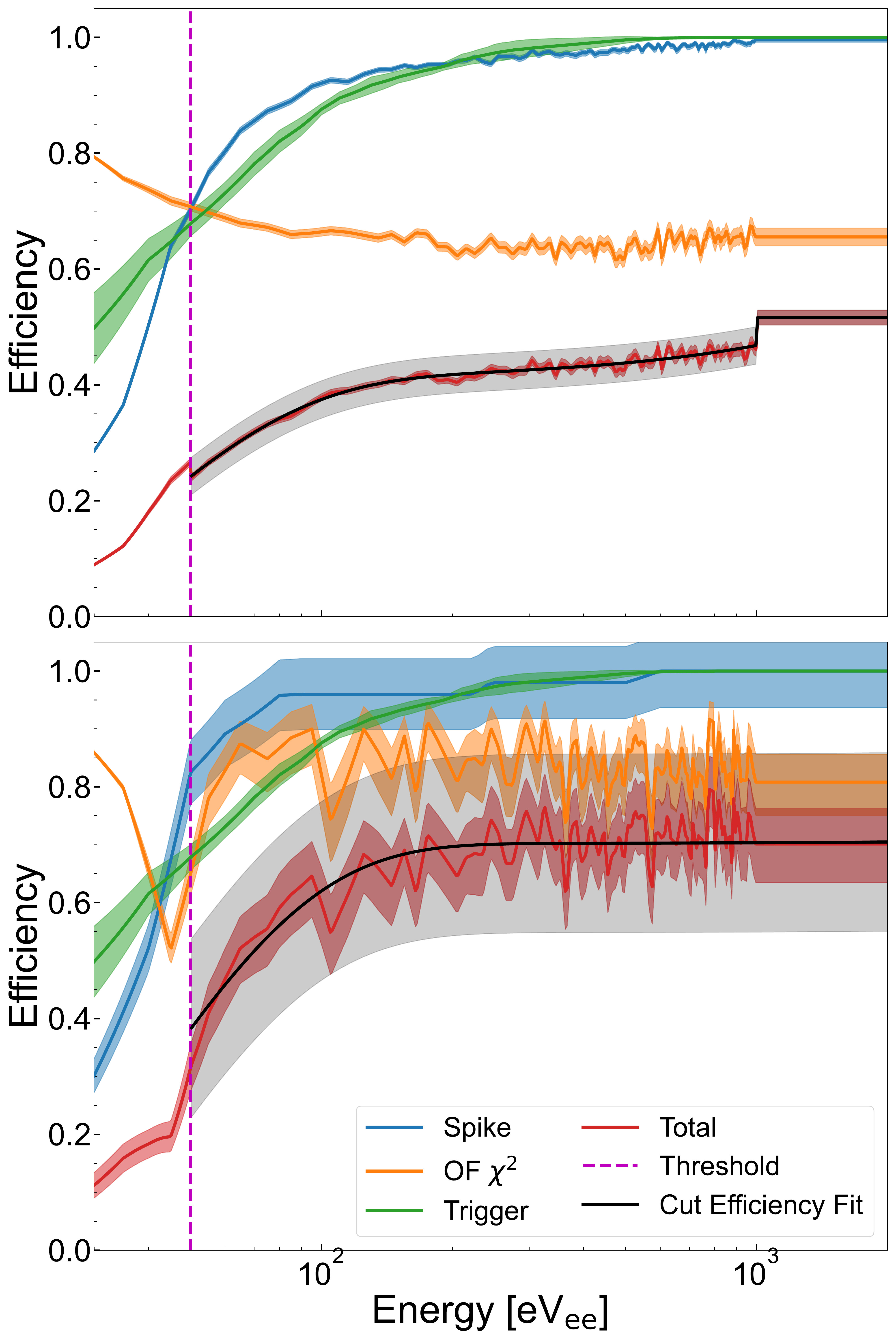}
    \caption{(Color online) Energy-dependent efficiencies for PuBe (upper) and background (lower) datasets. Black
curves show smooth functional forms of the total cut efficiency used for further analysis. 
} 
    \label{fig:cut_effs}
\end{figure}

The following quality cuts were applied to both background and PuBe data. The cut efficiency was
defined as the good event fraction at a given energy that survive the cut. The cut efficiency is
shown in Fig.~\ref{fig:cut_effs} for background and neutron datasets.  Further details on these
cuts have been given by Mast~\cite{Mast_thesis}. 

\noindent
\textbf{Baseline Cut: }Events were removed if the pre-pulse baseline average or variance was
excessively high. This removed events on the tail of an earlier pulse, as well as noisy data. The
efficiency was calculated from the passage fraction of randomly triggered traces and was found to
be 0.820$\pm$0.001 (energy independent).  

\noindent
\textbf{Pileup Cut: } Events containing multiple pulses were removed based on any of three
criteria: (1) the ratio between the integral of the trace (which includes all pulses present) and
the fitted OF amplitude (which fits a single pulse by definition) was larger by 3$\sigma$ than the
median value; (2) the OF delay was within 1\,$\mathrm{\mu s}$ of the early edge or 2\,$\mathrm{\mu
s}$ of the late edge of the 100 $\mathrm{\mu s}$-wide fitting window; and (3) the OF delay was
more than 70\,$\mathrm{\mu s}$ earlier or more than 10\,$\mathrm{\mu s}$ later than the 50\% point
on the rising edge computed using a pulse-shape characterization algorithm~\cite{Clarke2003}.  The
passage fraction of the pileup cut was 0.965$\pm$0.001 which we used as the energy-independent
efficiency for the cut. 

\noindent
\textbf{Spike Cut: }Events with unusual pulse shapes were occasionally observed, but easily
removed due to unnaturally-fast fall times. The efficiency was energy dependent and is shown in
Fig.~\ref{fig:cut_effs}. It was calculated as the passage fraction for datasets in run periods that
were mostly free of such events.  

\noindent
\textbf{OF $\chi^2$ Cut: }The goodness of the OF fit is quantified with a $\chi^2$ calculated in
frequency space and weighted by the average noise power spectral density. An energy-dependent cut
was defined to remove events with $\chi^2$ per degree of freedom values that exceeded 1.25.
Events removed by this cut were mostly ordinary pulses. As such, we assumed the passage fraction
of this cut to be its energy-dependent efficiency, shown in Fig.~\ref{fig:cut_effs}. 

\noindent
\textbf{Low Energy Trigger Burst Cut: } Short bursts of events below 150\,\eVee~were occasionally
observed.  The bursts comprised high rate (above 1\,kHz) periods of otherwise good pulses in the
space of tens to hundreds of ms.  Bursts were almost non-existent in background data but
significant in the two high-rate datasets. To identify events in bursts we examined the proximity
between consecutive low-energy (below 1\,keV) events in the event sequence. Low-energy events were
required to be sequentially separated by greater than 20 events. After applying this criterion the
resulting event sequence was consistent with a random distribution of low-energy events. The
background dataset was consistent with having no burst events, so this cut was not applied to it.

The cut only removed events between the 50\,\eVee~analysis threshold and 1\,\keVee, where the
resulting efficiency was 0.893$\pm$0.001 and the estimated leakage fraction was less than two
tenths of a percent. Outside of 50--1000\,\eVee, the efficiency was 1. 

\textbf{Trigger: } The DAQ system trigger efficiency was calculated using a distribution of
simulated pulses constructed from higher-energy events where the trigger efficiency was 100\%,
then scaled down to simulate lower energy events and added to noise from randomly-triggered
traces. The trigger algorithm was applied to this distribution to generate an efficiency curve as
a function of energy as shown in Fig.~\ref{fig:cut_effs}.

The DAQ has a limited speed, such that some events which trigger are not written.  The write
efficiency is energy-independent, but rate-dependent. A write efficiency of 0.617$\pm$0.004 and
0.815$\pm$0.004 for the PuBe and background data respectively was measured by comparing the rate
of pulses that should pass the trigger to the actual write rate. 

\FloatBarrier
\section{\label{sec:cap_spec}Capture Spectrum}
When a nucleus relaxes after neutron capture, it passes through a number of nuclear levels,
emitting as many gammas as levels visited. This de-excitation process is called a cascade and
typically it happens fast enough that all the dynamics appear in one measured event. The
properties of the resulting NR depend on the specific cascade realized in that event. Since this
is the signal we are attempting to extract from the neutron data, we carefully simulated the
cascade event and understand the resulting NR spectrum.

The energy deposits were modeled for individual cascades and then combined with the correct
probabilities to make the total spectrum~\cite{Villano2022}. Each probability is
derived from both the relative abundance~\cite{abundances} of the isotope and its capture
cross-section~\cite{doi:10.1080/18811248.2011.9711675}. The probabilities for each cascade are
inferred from the literature~\cite{PhysRevC.46.972}.

Modeling is simple for one-step cascades. For multi-step cascades, several parameters become
important, including the stopping properties of recoils, the half-lives of individual energy
levels, and the angular distribution of emitted gammas. For stopping properties, we used
constant-acceleration stopping equal to the average of the Lindhard stopping
power~\cite{osti_4701226}. Half-lives of intermediate levels were taken from measurements where
possible; otherwise, Weisskopf estimates were used~\cite{PhysRev.83.1073}. The angular
distribution of emitted gammas was taken as isotropic. For multi-step cascades the deposited
energy is not always a single value like it is for one-step cascades. Depending on the level
parameters multi-step cascades can give single values or broad spectra (if there is a
decay-in-flight for recoil atom). 

\begin{figure}
	\includegraphics[width=\columnwidth]{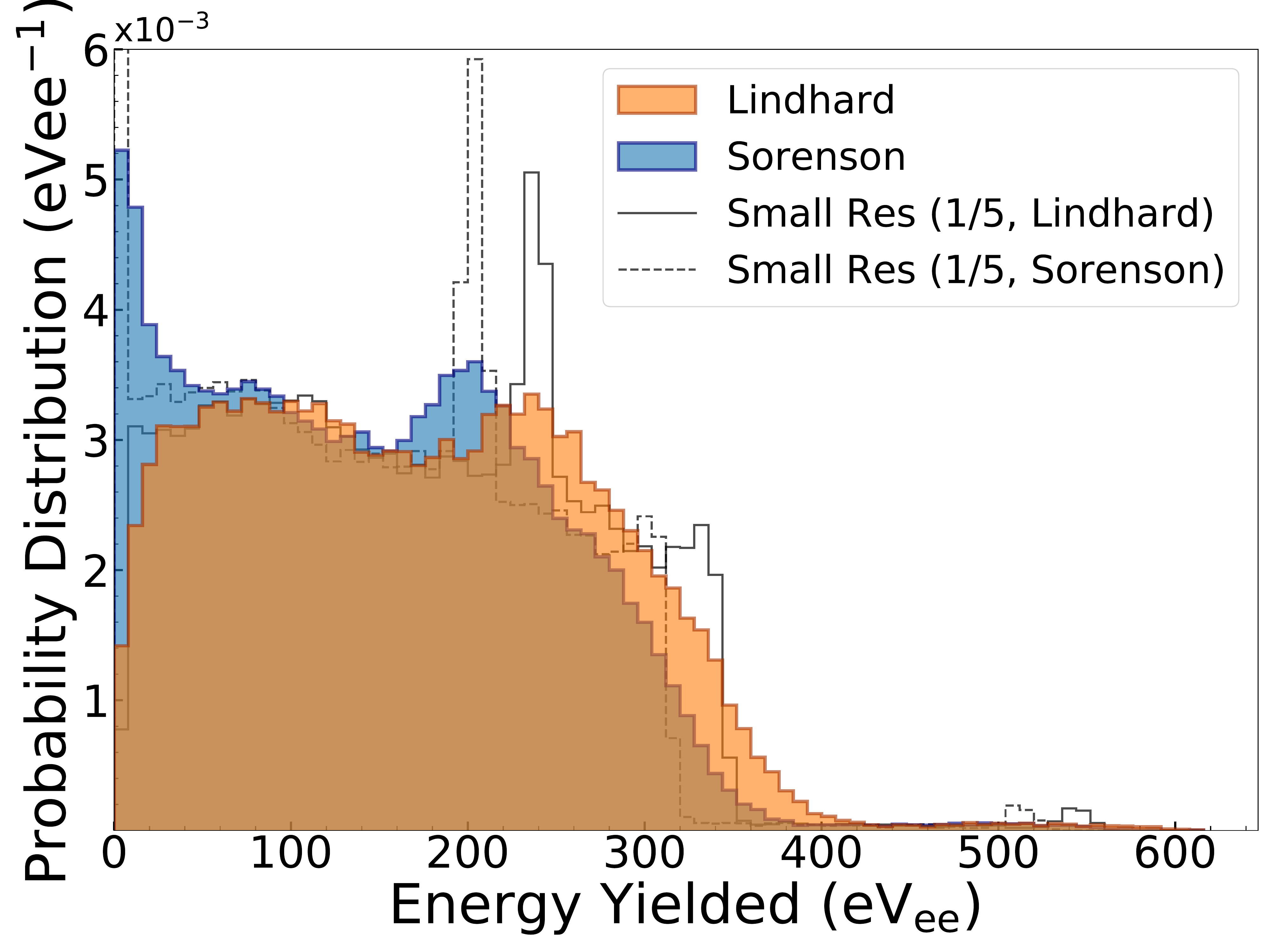}
	\caption{
		(Color online) Overlaid histograms comparing the yielded energy PDFs for
Sorensen\cite{PhysRevD.91.083509} and Lindhard\cite{osti_4701226} models, including the resolution
of the current detector.  The histograms are comprised of approximately $10^6$ simulated cascades.
The orange (front) filled histogram represents the Lindhard model while the blue (back) filled
histogram represents the Sorensen model.  In the Sorensen model, many points are pushed to zero
due to the presence of a cutoff energy, leading to a peak in the first bin that is not present in
the Lindhard model.  For both models, we use $k=0.178$ and $q=0.00075$.  The solid-line unfilled
histogram represents the Lindhard model with a 5x better resolution, and the
dashed unfilled histogram represents the Sorensen model with that 
resolution.  Both of these show much taller, narrower peaks than their counterparts.  }
	\label{fig:svl_overlay}
\end{figure}

Silicon has more than 80 such cascades and many have low probability.  While we did model all the
cascades, we only included the six most common cascades for $^{29}\mathrm{Si}$ (capture on
$^{28}\mathrm{Si}$) since they provided 94\% of the total spectrum for that isotope and adding in
all the cascades did not result in a significant change in the shape of the curve.  A similar
strategy applied to the other isotopes of silicon led to the selection of the four most common
cascades for $^{30}\mathrm{Si}$ and $^{31}\mathrm{Si}$ for a total of 14 cascades used. We assumed
natural abundances of isotopes. Table~\ref{tab:acc_prob} shows the parameters of all the cascades
included in our modeling. The expected distributions of the ionization energies due to these
capture events are shown in Fig.~\ref{fig:svl_overlay} for two different yield models. 
\begin{table*}[!hbt]
\begin{tabular}{ c  c  c  c  c  c }
\hline
\hline 
	Cascade ID (CID)  &  Isotope &  Prob. (\%)   & Energy Levels (keV) & Half-Lives (fs) & Cumulative Contribution (\%) (Lind./Sor.) \\ \hline
	1&	 $^{28}$Si&	 62.6&	4934.39         &   0.84       & 63.6/63.7 \\
	2&	 $^{28}$Si&	 10.7&	6380.58, 4840.34&	0.36, 3.5  & 75.0/74.0 \\
	3&	 $^{28}$Si&	 6.8 &	1273.37         &	291.0      & 83.3/83.4 \\
	4&	 $^{28}$Si&	 4.0 &	6380.58         &	0.36       & 88.1/88.7 \\
	5&	 $^{28}$Si&	 3.9 &	4934.39, 1273.37&	0.84, 291.0& 91.7/91.9 \\
	6&	 $^{28}$Si&	 2.1 &	-               &	-          & 94.3/94.8 \\
	7&	 $^{29}$Si&	 1.5 &	6744.1.0        &	14         & 96.1/96.7 \\
	8&	 $^{30}$Si&	 1.4 &	3532.9, 752.2.0 &	6.9, 530   & 97.1/97.3 \\
	9&	 $^{29}$Si&	 1.2 &	7507.8, 2235.3.0&	24, 215    & 98.5/98.5 \\
	10&	 $^{29}$Si&	 0.4 &	8163.2.0        &	w(E1)      & 99.0/99.1 \\
	11&	 $^{30}$Si&	 0.4 &	5281.4, 752.2.0 &	w(E1), 530 & 99.3/99.3 \\
	12&	 $^{29}$Si&	 0.3 &	-               &	-          & 99.7/99.8 \\
	13&	 $^{30}$Si&	 0.3 &	4382.4, 752.2.0 &	w(E1), 530 & 100./100. \\
	14&	 $^{30}$Si&	 0.0 &	-               &	-          & 100./100. \\
\hline
\hline
\end{tabular}
   \caption{\label{tab:acc_prob}
	A table displaying the cumulative fractional contribution of each Cascade Identifier (CID)
	for both the Lindhard and Sorensen models This table includes only the cascades used.  The
	statistics reported only include events which were above the detector threshold.  The
	isotope listed is the isotope on which the neutron captures; the energy levels and
	half-lives are therefore for an isotope of silicon with one more neutron.  A half-life
	entry of w(E1) specifies that the half-life is unknown and the Weisskopf estimate for an
	electric dipole transition was used~\cite{PhysRev.83.1073}.  
}
\end{table*}

Our model accounted for how often the gammas from capture exited the detector without depositing
energy, leaving only the isolated NR behind.  This selection did not distort the spectral shape
much, cutting out roughly 10\% of all the events.  These models were used to simulate $10^6$
capture cascades for comparison to the observed data (see Sec.~\ref{sec:fit}).

\FloatBarrier
\section{\label{sec:sim}Non-capture spectra}
An ideal neutron source would produce only thermal neutrons, but a PuBe source also produces gamma
radiation and higher-energy neutrons which reach the detector and produce elastic recoils, some of
which deposit energies in the analysis region. For a full analysis it was necessary to model these
other components of the observed spectrum. 

For these non-capture events (ERs and non-capture NRs), we directly used the deposited recoil
energies from ER and NR hits as modelled in \textit{Geant4}~\cite{1610988,AGOSTINELLI2003250}. We
used the version \textit{Geant4.10.1.p02}. A complete model of the laboratory configuration was
used, including the PuBe source and housing, all shielding elements, the refrigerator frame and
main refrigerator components, the main components of the hardware supporting the detector, and the
floor, ceiling and walls, with the intent to fully account for complex neutron paths.  We base our
PuBe simulated source spectra on Ref.~\cite{PhysRev.98.740}.  

In the \textit{Geant4} simulation, high-precision electromagnetic physics and neutron physics were
used~\cite{BROWN201477,THULLIEZ2022166187}. Although newer models of \textit{Geant4} (after
\textit{Geant4.10.5}) use an upgraded coherent $\gamma$-nucleus scattering~\cite{OMER201743}, our
simulation uses the older EPDL model~\cite{osti_295438}. The EPDL model has significantly
different angular distributions -- although the total cross sections are close -- and could affect our
simulation. The difference is unlikely to change our results because the $\gamma$ environment is
dominated by capture $\gamma$'s from the surrounding materials and across a wide range of energies
(1--10\,MeV). That spectrum cannot create features similar to capture-induced NRs. 

Direct neutron scatters (single or multiple) are also not likely to change our results.
Hi-precision neutron physics \textit{NeutronHP} is included in our \textit{Geant4} physics list.
While the modeling is not likely to be perfect our direct-scatter neutron environment has a wide
range of energies--from below 1\,MeV to as high as around 9\,MeV with no strong energy features.
This neutron spectrum will create a nearly featureless quasi-exponential background with very
little impact from multiple scatters~\cite{SuperCDMS:2022nlc}.    

Recently, there has been interest in inelastic processes that can occur at these energies, namely
the Migdal effect and atomic
bremsstrahlung~\cite{Ibe:2017yqa,PhysRevLett.121.101801,PhysRevLett.118.031803}. We did not model
these backgrounds because a calculation showed that they would be 2--4\% of the expected capture
signal. 


\FloatBarrier
\section{\label{sec:fit}Fitting}
Our data analysis consists of fitting a simulated PuBe spectrum to background-subtracted data.
The simulated spectrum consists of both thermal neutron-capture events 
as well as PuBe-generated non-capture ERs and NRs.  Data from the background dataset is normalized and subtracted from the PuBe signal data before being compared to
simulation. We accounted for the data-taking and cut efficiencies by applying the relevant
corrections to the data after the cuts.

\textbf{Integral method:} Our preferred method of constraining the ionization yield is to fit a
well-motivated theoretical model with a small number of parameters to the data. However, the
ionization yield as a function of energy has been shown to be a poorly-understood theoretical
construct~\cite{PhysRevD.103.122003,PhysRevD.94.082007,PhysRevD.94.122003}.  Our approach to deal
with this situation was to use an ``integral method'' similar to
Chavarria~\cite{PhysRevD.94.082007} while assuming consistency with the higher-energy Izraeliavitch
data~\cite{Izraelevitch_2017}. We did this with and without the inclusion of the neutron-capture
induced NRs to give a generic understanding of the plausibility of the ionization functions in
each situation. Note that the integral method can reproduce given experimental data with
\emph{any} NR component since it essentially has infinitely many degrees of freedom.  This
procedure was developed in detail by Mast~\cite{Mast_thesis}. 

\begin{figure*}[h]
    \centering
    \includegraphics[width=2\columnwidth]{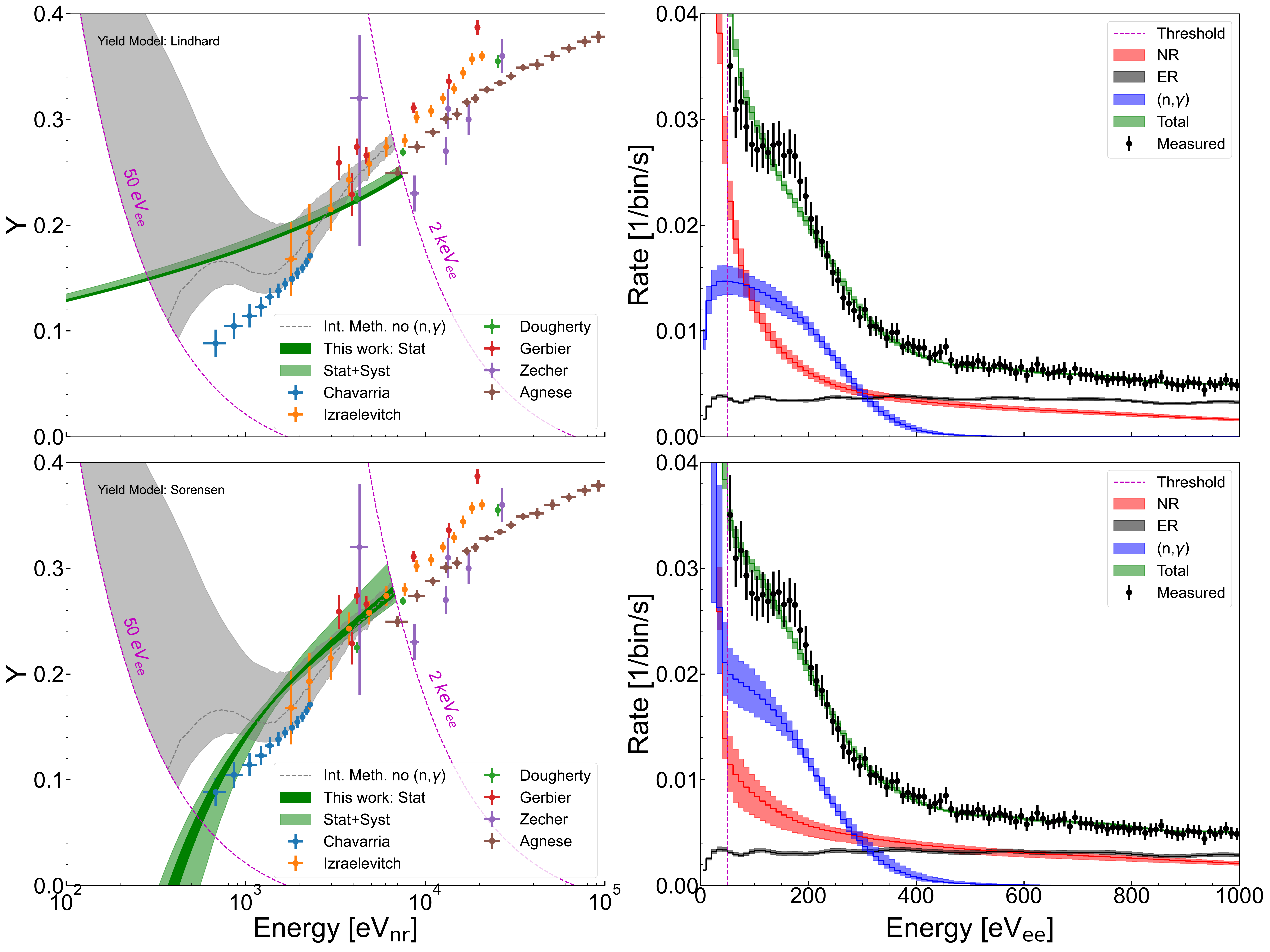}
	\caption{(Color online) MCMC fitting results for Lindhard (top) and Sorensen (bottom) yield models. Left:
Best-fit yield curves using the specified yield model are shown with statistical and systematic
errors, in comparison with multiple published measurements. The detector threshold level is shown
as the low magenta dashed curve. Right: Range of best-fit background-subtracted reconstructed
spectra. Shaded bands represent the 1-sigma equivalent range of rates in each energy bin.  On each
of the left-hand plots the result of the integral method \emph{without} including the
neutron-capture induced NRs is shown (grey dashed lines and grey bands); the resulting ionization
yield function is poorly constrained and oddly shaped implying the necessity of the
neutron-capture induced NR contribution.}
    \label{fig:mcmc_fits}
\end{figure*}

Executing the procedure assuming that neutron-capture induced NRs were \emph{not} present produced
an oddly shaped yield curve with an anomalous increase below 1\,keV recoil energy (see
Fig.~\ref{fig:mcmc_fits}). Conversely, including the neutron-capture induced NRs (not shown) gave
a yield curve that was better behaved at low energies and more consistent with previous
measurements, especially those of the DAMIC collaboration~\cite{PhysRevD.94.082007}. 

The results were calculated with the assumption that the Fano factor for NRs, $F_{NR}$=0.1161, is
the same as for ERs. Repeating this exercise with different values showed almost no change in the
resulting yield band for $F_{NR}<5$. This is not surprising because even at a low NR Fano factor,
the features in the capture spectrum are smeared. 

\textbf{Markov Chain Monte Carlo (MCMC) method:} A direct fit to the data was also performed,
including different parameterized yield models.  As mentioned above, no ionization yield model in
the literature seems fully appropriate for NRs at these low recoil energies, but performing the
fit allowed us to compare our data using well-established statistical techniques and ionization
yield models with a limited number of parameters. We avoided optimizing our fits to an arbitrary
functional form without convincing theoretical motivation. Nonetheless, our results imply: (1) a
clear identification of the neutron-capture induced NR signal; (2) a further indication that the
long-popular Lindhard model~\cite{osti_4701226} is not a complete description; and (3) a
preference for a yield model which goes to near-zero ionization yield at a finite recoil energy --
a possibility with far-reaching implications for DM or CE$\nu$NS science.

The fitting was accomplished using the Markov chain Monte Carlo (MCMC) ensemble sampler emcee
\cite{Foreman_Mackey_2013}. Our method follows closely the method of
Scholz~\cite{PhysRevD.94.122003} and was developed by Mast~\cite{Mast_thesis}. The fit was
performed with the following yield models ($Y(E_r)$): Lindhard~\cite{osti_4701226},
Sorensen~\cite{PhysRevD.91.083509}, Chavarria~\cite{PhysRevD.94.082007}, and Adiabatic Correction
(AC)~\cite{PhysRevD.36.311}. Independent scaling factors for each of the three simulated spectra
-- capture, ER, and non-capture NR -- were included as fit parameters. The Fano factor for NRs was
also allowed to float in the fits.  To obtain the posterior distributions via the MCMC technique,
a flat prior distribution in reasonable parameter ranges was assumed. 

To accommodate the asymmetric uncertainties which resulted from our cuts and background
subtraction methodology, we described the observed counts in each bin with a
Split-Normal~\cite{10.1214/13-STS417} distribution with upper and lower uncertainties
$\sigma_{\mathrm{hi}}$ and $\sigma_{\mathrm{low}}$ respectively.  The log likelihood function is
\begin{widetext}
\begin{equation} 
    \ln(\mathcal{L}_{SNorm}(\vec{c};\vec{\mu},\vec{\sigma}_{low},\vec{\sigma}_{hi})) = \sum_i \left[ \frac{1}{2}\ln\left(\frac{2}{\pi}\right) - \ln(\sigma_{low,i}+\sigma_{hi,i}) -\frac{1}{2}\left(\frac{c_i-\mu_i}{\sigma_i}\right)^2\right]
\end{equation}
\end{widetext}
where $\vec{c}$ -- which implicitly depends on all the fit parameters -- is the set of simulated
counts, $\vec{\mu}$ are the average measured rates for each bin,
$\vec{\sigma}_{hi}~(\vec{\sigma}_{low})$ is the width parameter for points above (below) $\mu$,
and $\sigma_i$ is a piecewise function giving the upper width if $c_i$ is above $\mu_i$ and the
lower width otherwise.

\begin{table}[h]
    \centering
    \begin{tabular}{| c || c | c | c | }
        \hline
        Model & $\chi^2$/DOF & $\chi^2$/DOF (no cap.) & par.\\
        \hline\hline
        Lindhard & 722.8/190=3.804 & 1653.7/191=8.659 & $k$ \\ 
        \hline
        Sorensen & 306.7/189=1.623 & 1765.7/190=9.293 & $k$,$q$\\
        \hline
        Chavarria &  670.4/189=3.547 & 2010.3/190=10.581 & $k$,$a$ 	\\
        \hline
        AC & 525.0/189=2.778 & 1808.4/190=9.518 & $k$,$\xi$ \\
        \hline
    \end{tabular}
    \caption{Table of best-fit $\chi^2$/DOF values for several yield models. Calculated using only statistical uncertainties.}
    \label{tab:chisq}
\end{table}

The maximum likelihood goodness-of-fits are shown in Tab.~\ref{tab:chisq}. While the Sorensen
model yields the best fit by far, even that fit is not particularly good. The table also shows the
goodness-of-fits for a parallel fit that does \emph{not} include the neutron-capture component. In
all cases there is a strong preference for the inclusion of the neutron-capture component (at
least $\sim$25$\sigma$). Using a Likelihood Ratio Test~\cite{Wilks:1938dza,PhysRevD.99.062001}
there is a preference to reject the Lindhard model in favor of any other with a p-value of at most
4.4$\times$10$^{\mathrm{-13}}$. The best fit values of the scaling factors of the simulated
spectra indicate that the absolute rates predicted by the \textit{Geant4} simulation do not agree with the
observed data. 

\textbf{MCMC results:} The models we believe deserve the most focus are the Lindhard model for
historical significance and the Sorensen model because of its yield falloff at low energies. The
Sorensen model is characterized by
$Y_{\mathrm{Sor}}(E_r,k,q)=Y_{\mathrm{L}}(E_r,k)-q/\varepsilon(E_r)$, where $Y_{\mathrm{L}}$ is
the Lindhard model and $\varepsilon(E_r)$ is the unitless version of $E_r$ used in the Lindhard
Model. The Sorensen model was the best fitting model and the parameters were
$k$\,=\,0.151$^{+040}_{-0.007}$ and $q$\,=\,1.96$^{+1.32}_{-0.54}\times$10$^{\mathrm{-3}}$.
Figure~\ref{fig:mcmc_fits} shows the details of the fit results.  

\section{\label{sec:conclusions}Conclusions}

This experiment studied nuclear recoils in silicon down to 450\,eV and analyzed the spectrum to
find evidence for induced NR via neutron capture. The final measured spectrum strongly prefers a
thermal neutron capture-induced NR component, a preference that corresponds to 25$\sigma$ or more
for each ionization model studied.  If the (n,$\gamma$) process is \emph{not} included, the
resulting shape of the ionization yield function $Y(E_r)$ becomes unusually distorted to make up
for it, as demonstrated by the integral method. 

Our results favor the Sorensen ionization yield model (which has a low-energy ionization cutoff)
to the standard Lindhard model, giving a p-value of the Likelihood Ratio
Test~\cite{PhysRevD.99.062001} of less than 4.4$\times$10$^{-13}$.  While it is clear that a
``perfect'' fit to the data is possible for \emph{some} ionization yield model $Y(E_r)$, none of
the proposed models fit well. The ``perfect'' fit that would be given by the integral method with
(n,$\gamma$) included provides little understanding of the process and in particular how it may
depend on field strength and/or temperature, so it was omitted. More theoretical work on this
process is required first.   

Neutron-capture induced events provide an excellent window into very low-energy NRs. Further
studies on NR ionization yield in silicon are necessary to establish the behavior at low energies
and to quantify a possible 100\,eV-scale yield threshold.  Improvements are planned for the next
experiment. The recoil energy threshold can be lowered by better background mitigation and
improved detector resolution. Tagging the gammas emitted after capture will be a significant
improvement in the experimental technique, since the capture spectrum can then be isolated.

The data that support the findings of this study are made openly available with the Open Science
Framework (OSF)~\cite{Harris_Villano_Fritts_2021}. 

\begin{acknowledgements}
The authors would like to thank the SuperCDMS collaboration for the use of the detector and
readout electronics. This work was supported by DOE grants DE-SC0012294, DE-SC0021364, and grant
NSF-1743790 via the Partnerships for International Research and Education Program (PIRE), and the
Germanium Materials and Detectors Advancement Research Consortium (GEMADARC).
\end{acknowledgements}

\clearpage
\bibliography{prdrefs_short.bib}
\bibliographystyle{apsrev4-2}

\end{document}